\documentclass[12pt]{article}
\usepackage{amsfonts}
\newcommand{\E}{{\rm e}} 
\newcommand{\I}{{\rm i}} 
\newcommand{\e}{{\mathbb E}\,}
\makeatletter
\begin{document}
\title{\bf\LARGE Coherent-state quantization\\ of constrained fermion systems}
\author{Georg Junker\\[2mm] Institut f\"ur Theoretische Physik, Universit\"at
Erlangen-N\"urnberg,\\ Staudtstr.\ 7, D-91058 Erlangen, Germany\\[2mm]
and\\[2mm]
John R.\ Klauder\\[2mm]
Departments of Physics and Mathematics, University of Florida,\\
Gainesville, FL-32611, USA}
\maketitle
\abstract{The quantization of systems with first- and second-class constraints
within the coherent-state path-integral approach is extended to quantum systems
with fermionic degrees of freedom. As in the bosonic case the importance of
path-integral measures for Lagrange multipliers, which in this case are in
general expected to be elements of a Grassmann algebra, is explicated.
Several examples with first- and second-class constraints are discussed. }
\section{Introduction}  
The quantization of constrained systems has recently been reexamined  
\cite{Klauder97,Kla97a,Kla97b,Kla97c} from the point of view of 
coherent-state path integrals,  
which revealed significant differences from the standard operator and  
path-integral approaches. The aim of this contribution is to extend this   
approach, formulated for bosonic degrees of freedom,  
to fermionic systems. That is, we will discuss the generalization  
of the approach of \cite{Klauder97} to constrained quantum systems with   
fermionic degrees of freedom. As in the bosonic case we will utilize the   
(fermion) coherent-state path-integral approach. In essence the basic   
idea of inserting projection operators via proper path-integral measures   
for Lagrange multipliers is the same as in the bosonic case   
\cite{Klauder97}. Therefore, we will closely follow the approach   
of \cite{Klauder97} and put more emphasize on the presentation of   
various examples with first- as well as second-class constraints. We   
will omit a discussion of the classical version of such systems, that   
is, the so-called pseudomechanics \cite{Cas76c,HenTei92} which is the   
classical dynamics of Grassmann degrees of freedom. Also the   
quantization of such systems (without constraints) is well discussed in   
the literature \cite{Cas76b,HenTei92}. Note that due to the Grassmannian   
nature, the classical dynamics formulated in phase-space always exhibits   
second-class constraints which, however, can easily be removed   
\cite{HenTei92}. For these reasons we will exclusively concentrate our   
attention on fermionic quantum systems with operator-valued constraints.   
  
The outline of this paper is as follows. In Sect.\ 2 we will review some  
basic concepts of quantum systems consisting of $N$ fermionic degrees of   
freedom. In particular, we discuss several properties of fermion   
coherent states and the associated path-integral approach. In doing so   
we shall also give a minimal review Grassmann theory. Section 3 is   
devoted to a general discussion of first-class constraints including a   
construction method for projection operators following \cite{Klauder97}.  
In Sect.\ 4 several examples with first-class constraints are   
discussed. In Sect.\ 5 we briefly outline the generalization of the   
treatment of second-class constraints of \cite{Klauder97} to fermion   
systems. Section 6 presents a discussion for a wide range of odd 
second-class   
constraints on the basis of typical examples. Finally, in Sect.\ 7 we   
consider an example of a constrained boson-fermion system.  
  
\section{Basic concepts of fermionic degrees of freedom}   
\subsection{Grassmann numbers}  
It is well-known that Grassmann numbers may serve as   
classical analogues of fermionic degrees of freedom. To be more explicit,   
the ``classical phase space'' of $N$ fermions may be identified with the   
Grassmann algebra ${\Bbb C}B_{2N}$ over the field of complex numbers   
\cite{Corn89,ConGro94}, which is generated by the set   
$\{\bar{\psi}_1,\ldots,\bar{\psi}_N,\psi_1,\ldots,\psi_N\}$ of $2N$   
independent Grassmann numbers obeying the anticommutation relations   
\begin{equation}   
\begin{array}{l}
\{\psi_i,\psi_j\}:=\psi_i\psi_j+\psi_j\psi_i=0\;,\\[2mm]
\{\psi_i,\bar{\psi}_j\}=0\;,\quad \{\bar{\psi}_i,\bar{\psi}_j\}=0\;. 
\end{array}
\end{equation}   
This algebra allows for a natural ${\Bbb Z}_2$ grading by appointing   
a degree (also called Grassmann parity) to all homogeneous elements   
(monomials) of ${\Bbb C}B_{2N}$:   
\begin{equation}   
\deg\left(\bar{\psi}_{j_1}\cdots\bar{\psi}_{j_m}  
\psi_{i_1}\cdots\psi_{i_n}\right):=   
\left\{ \begin{array}{lcl} 0 & \mbox{ for }& m+n\mbox{ even}\\ 1 &   
\mbox{ for }& m+n\mbox{ odd} \end{array}\right. \;.  
\label{2} 
\end{equation}   
In other words, the even elements of ${\Bbb C}B_{2N}$ are commuting   
and the odd elements are anticommuting numbers. For further details we   
refer to the textbooks by Cornwell \cite{Corn89} and Constantinescu and   
de Groote \cite{ConGro94}. Here we close by giving the convention of   
Grassmann integration and differentiation used in this paper:   
\begin{equation}   
\int d\psi \,1=0\;,\quad   
\int d\psi\,\psi =1\;,\quad   
\frac{d}{d\psi }\,1=0\;,\quad   
\frac{d}{d\psi }\,\psi =1\;.  
\end{equation}   
Here $\psi$ stands for any of the $2N$ generators of ${\Bbb C}B_{2N}$,   
and the integration and differentiation operators are treated like odd   
Grassmann quantities according to the ${\Bbb Z}_2$ grading (\ref{2}).  
  
\subsection{Fermion coherent states}  
Throughout this paper we will consider quantum systems with a finite number,   
say $N$, of fermionic degrees of freedom which are characterized by   
annihilation and creation operators $f_i$ and $f_i^{\dagger}$, 
$i=1,2,\ldots,N$, obeying the canonical anticommutation relations  
\begin{equation}  
\{f_i,f_j\}=0\;,\quad  
\{f^{\dagger}_i,f^{\dagger}_j\}=0\;,\quad  
\{f^{\dagger}_i,f_j\}=\delta_{ij}\;.  
\end{equation}  
The corresponding Hilbert space is the $N$-fold tensor product of the 
two-dimensional Hilbert spaces ${\cal H}_i\equiv{\Bbb C}^2$ for a 
single degree of freedom,  
\begin{equation}  
{\cal H}={\cal H}_1\otimes{\cal H}_2\otimes\cdots\otimes  
{\cal H}_N={\Bbb C}^{2^{N}}\;.  
\end{equation}
A standard basis in this ``$N$-fermion'' Hilbert space ${\cal H}$ is the 
simultaneous eigenbasis of the number operators $f_i^{\dagger}f_i$:  
\begin{equation}  
f_i^{\dagger}f_i|n_1 n_2 \ldots n_N\rangle=n_i |n_1 n_2 \ldots n_N\rangle 
\;,\quad n_i=0,1\;,  
\end{equation}  
where   
\begin{equation}  
|n_1 n_2 \ldots n_N\rangle:= |n_1\rangle_1\otimes|n_2\rangle_2\otimes\cdots  
\otimes |n_N\rangle_N  
\end{equation}  
with $|n\rangle_i$ being a vector in the one-fermion Hilbert space 
${\cal H}_i$ on which the operators $f_i$ and $f_i^{\dagger}$ are acting via  
\begin{equation}  
\begin{array}{l}
f_i|0\rangle_i=0\;,\quad f_i|1\rangle_i=|0\rangle_i\;,\\[2mm]  
f_i^{\dagger}|0\rangle_i=|1\rangle_i\;,\quad f_i^{\dagger}|1\rangle_i=0\;.  
\end{array}
\end{equation}  
  
Fermion coherent states are defined in analogy to the canonical    
(boson) coherent states \cite{Mar59,OhnKas78,KlaSka85}. They qualitatively
differ, however,   
from the latter as the basic quantities labeling   
these states are not ordinary c-numbers but rather are odd Grassmann   
numbers. To be more precise, they are the generators of the classical   
phase space ${\Bbb C}B_{2N}$. For simplicity let us consider in the   
following discussion only one fermionic degree of freedom, that is, we set   
$N=1$ and subscripts will be omitted. Then the fermion coherent states   
are defined \cite{Mar59,OhnKas78,KlaSka85} as follows:   
\begin{equation}  
|\psi \rangle:=  
\exp\{-\textstyle\frac{1}{2}\bar{\psi }\psi \} \E^{f^{\dagger}\psi }|0\rangle  
=\exp\{-\textstyle\frac{1}{2}\bar{\psi }\psi \}  
\Bigl(|0\rangle-\psi |1\rangle\Bigr).  
\label{9} 
\end{equation}  
The corresponding adjoint states read  
\begin{equation}  
\langle \psi |:=  
\exp\{-\textstyle\frac{1}{2}\bar{\psi }\psi \} \langle 0|\E^{\bar{\psi} f} =  
\exp\{-\textstyle\frac{1}{2}\bar{\psi }\psi \}  
\Bigl(\langle 0|+\bar{\psi}\langle 1|\Bigr).  
\end{equation}  
The normalized states (\ref{9}) form an overcomplete set in the one-fermion 
Hilbert space ${\Bbb C}^2$, that is,  
\begin{equation}  
\begin{array}{rcl}  
\langle \psi _{1}|\psi _{2}\rangle &=&   
\exp\{-\textstyle\frac{1}{2}\bar{\psi }_{1}\psi _{1}\}   
\exp\{-\textstyle\frac{1}{2}\bar{\psi }_{2}\psi _{2}\}   
\exp\{\bar{\psi }_{1}\psi _{2}\} \\[2mm]  
&=&\exp\{-\frac{1}{2}\bar{\psi }_{1}(\psi _{1}-\psi _{2})+  
       \frac{1}{2}(\bar{\psi }_{1}-\bar{\psi }_{2})\psi _{2}\}  
\end{array}  
\label{10}
\end{equation}  
and provide a resolution of the identity ${\bf 1}$ via  
\begin{equation}  
\begin{array}{l}  
\displaystyle  
\int d\bar{\psi }d\psi\, |\psi \rangle\langle \psi |\\[2mm]
\quad =\displaystyle  
\int d\bar{\psi }d\psi\Bigl[|0\rangle\langle 0|-\psi |1\rangle\langle 0|+  
\bar{\psi }|0\rangle\langle 1|-\bar{\psi }\psi {\bf 1}\Bigr]\\[2mm]  
\quad =\displaystyle  
\int d\bar{\psi }d\psi\Bigl[|0\rangle\langle 0|+|1\rangle\langle 0|\psi+  
\bar{\psi }|0\rangle\langle 1|+\psi \bar{\psi }{\bf 1}\Bigr]={\bf 1}\;.  
\end{array}  
\label{12} 
\end{equation}  
In the above we have already made use of a ${\Bbb Z}_2$ grading in analogy  
to that of Grassmann numbers. That is, we have appointed even and odd 
Grassmann degrees to the fermion coherent states and the operators 
\cite{OhnKas78}:  
\begin{equation}
\begin{array}{l}  
\deg(|0\rangle)=\deg(|\psi\rangle)=\deg(\langle\psi|)=0\;,\\[2mm]
\deg(|1\rangle)=\deg(f)=\deg(f^{\dagger})=1\;,  
\end{array}
\end{equation}
from which follow rules like  
\begin{equation}   
\psi |0\rangle=|0\rangle\psi \;,\quad\psi |1\rangle=-|1\rangle\psi\;,    
\quad\psi f=-f\psi\;,\quad \mbox{etc.}  
\end{equation}  
Finally, we mention that the fermion coherent states are eigenstates of  
the annihilation and creation operators  
\begin{equation}   
f|\psi \rangle=\psi |\psi \rangle=|\psi \rangle\psi \;,\quad   
\langle \psi |f^{\dagger}=\bar{\psi }\langle \psi |=\langle \psi |\bar{\psi }
\end{equation}  
and as a consequence the coherent-state matrix element of a normal-ordered  
operator $G(f^{\dagger},f)=\,:\!G(f^{\dagger},f)\!:$ reads  
\begin{equation}  
\langle \psi _{1}|G(f^{\dagger},f)|\psi _{2}\rangle=   
G(\bar{\psi }_{1},\psi _{2})\langle \psi _{1}|\psi _{2}\rangle\;.  
\end{equation}  
All of the above properties can trivially be generalized to the case of $N>1$  
degrees of freedom. In this case the fermion coherent states are essentially  
the ordered direct product of $N$ one-fermion coherent states \cite{KlaSka85}.
For example, in the case of two degrees of freedom these fermion   
coherent states read  
\begin{equation}  
\begin{array}{rl}  
\displaystyle  
|\Psi \rangle & :=|\psi _{1}\rangle\otimes|\psi _{2}\rangle\\[2mm]
& =\E^{-\bar{\Psi }\cdot\Psi/2}  
\Bigl(|00\rangle+|10\rangle\psi _{1}+|01\rangle\psi _{2}-|11\rangle  
\psi_{1}\psi_{2} \Bigr)\;,\\[3mm]  
\langle\Psi | & :=\langle\psi _{1}|\otimes\langle\psi _{2}|\\[2mm]
& =\E^{-\bar{\Psi }\cdot\Psi/2}  
\Bigl(\langle00|+\bar{\psi}_{1}\langle 10|+\bar{\psi }_{2}\langle 01|-  
\bar{\psi}_{1}\bar{\psi}_{2} \langle 11|\Bigr)\;,  
\end{array}  
\label{17} 
\end{equation}  
where we have set  
$\bar{\Psi}\cdot\Psi  :=\bar{\psi}_{1}\psi_{1}+\bar{\psi}_{2}\psi_{2}$.  
This notation naturally generalizes to cases with even more fermions, 
for example, 
\begin{equation} 
\langle\Psi''|\Psi'\rangle=\E^{-\bar{\Psi}''\cdot\Psi''/2}\,
\E^{-\bar{\Psi}'\cdot\Psi'/2}\,\E^{\bar{\Psi}''\cdot\Psi'}\;, 
\end{equation} 
and we will adopt this obvious generalization throughout this paper.  
  
\subsection{Fermion coherent-state path integrals}  
As in the standard canonical case one can represent the fermion-coherent-state
matrix element of the time-evolution operator $\exp\{-\I tH\}$ in terms of a   
coherent-state path integral \cite{Mar59,OhnKas78,KlaSka85}.   
For convenience we again consider a quantum   
system with a single degree of freedom which is completely characterized by 
an even normal-ordered Hamiltonian 
$H=H(f^{\dagger},f)=\,:\!\!H(f^{\dagger},f)\!\!:\,$.   
Hence, the coherent-state matrix element of the   
evolution operator (or propagator) is given by  
\begin{equation}  
\langle \psi''|\E^{-\I tH}|\psi'\rangle=  
\langle \psi''|\E^{-\I \varepsilon H}\E^{-\I \varepsilon H}\cdots  
\E^{-\I \varepsilon H}  
|\psi'\rangle  
\end{equation}  
where $\varepsilon :=t/N$. Inserting the completeness relation 
(\ref{12})  $N-1$ times  and taking the limit $\varepsilon\to 0$, that is 
$N\to\infty$ 
such that $N\varepsilon=t=const.$, one obtains the time-lattice definition   
($\psi_N:=\psi''$, $\psi_0:=\psi'$, $\Delta \psi _{n}:=\psi _{n}-\psi _{n-1}$,
$\Delta \bar{\psi} _{n}:=\bar{\psi }_{n}-\bar{\psi} _{n-1}$)  
\begin{equation}  
\begin{array}{l}  
\langle \psi''|\E^{-\I tH}|\psi'\rangle\\[2mm]
\quad=\displaystyle\lim_{\varepsilon\to 0} 
\prod_{n=1}^{N-1}\int d\bar{\psi }_{n}d\psi _{n}\prod_{n=1}^{N}
\langle \psi _{n}|\E^{-\I \varepsilon H}|\psi _{n-1}\rangle\\[2mm]
\quad=\displaystyle\lim_{\varepsilon\to 0}  
\prod_{n=1}^{N-1}\int d\bar{\psi }_{n}d\psi _{n}   
\prod_{n=1}^{N}\langle \psi _{n}|[1-\I \varepsilon H]|\psi _{n-1}\rangle\\[2mm]
\quad=\displaystyle\lim_{\varepsilon\to 0}  
\prod_{n=1}^{N-1}\int d\bar{\psi }_{n}d\psi _{n}   
\prod_{n=1}^{N} \E^{-\I \varepsilon H(\bar{\psi }_{n},\psi _{n-1})}   
\langle \psi _{n}|\psi _{n-1}\rangle\\[2mm]  
\quad=\displaystyle\lim_{\varepsilon\to 0}  
\prod_{n=1}^{N-1}\int d\bar{\psi }_{n}d\psi _{n}
\prod_{n=1}^{N}\exp\left\{-\frac{1}{2}\,\bar{\psi}_{n}\Delta\psi _{n}\right.\\
\qquad\displaystyle \left.
+\frac{1}{2}\,\Delta \bar{\psi }_{n}\psi _{n-1}-\I \varepsilon    
H(\bar{\psi }_{n},\psi _{n-1})\right\}  
\end{array}  
\end{equation}  
for the formal coherent-state path-integral representation of the propagator  
\begin{equation}
\begin{array}{l}
\displaystyle
\langle\psi''|\E^{-\I Ht}|\psi'\rangle=
\int{\cal D}\bar{\psi }{\cal D}\psi\\[2mm]
\qquad\displaystyle\times
\exp\left\{\I\int_{0}^{t}d\tau\textstyle  
\left[\frac{\I}{2}\left(\bar{\psi }\dot{\psi }-
\dot{\bar{\psi }}\psi \right)-H(\bar{\psi },\psi ) \right]  
\right\}\;.  
\end{array}
\end{equation}  
Similar path-integral expressions may also be derived for other matrix   
elements of the time-evolution operator \cite{OhnKas78,KlaSka85,EzaKla85}.
The above path-integral formulation is easily extended to several fermionic   
\cite{OhnKas78} and additional bosonic degrees of freedom   
\cite{EzaKla85}.   
  
The aim of this paper is to find similar path-integral representations of  
fermion systems subjected to additional constraints. In doing so we will 
closely follow the idea of \cite{Klauder97}, which incorporates proper  
projection operators via some additional path-integral measure for the  
Lagrange multipliers.  
  
\section{First-class constraints}  
The quantum systems under consideration are characterized by an even   
self-adjoint and normal-ordered Hamiltonian \linebreak
$H(f^{\dagger},f)$, where   
$f^{\dagger}$ and $f$ stand for the set $\{f^{\dagger}_1,\ldots,   
f^{\dagger}_N\}$ and $\{f_1,\ldots,f_N\}$, respectively. The quantum   
dynamics generated by this Hamiltonian is assumed to be subjected to   
constraints characterized by operator-valued normal-ordered functions of   
the fermionic annihilation and creation operators. Furthermore, we   
assume that these constraints have a well-defined Grassmann parity. Then, 
in the general case, we have two sets of constraints. One consists   
of even operators denoted by   
\begin{equation}  
\Phi_a\equiv\Phi_a(f^{\dagger},f)=\,\,:\!\Phi_a(f^{\dagger},f)\!:\,\,=  
\Phi_a^{\dagger}\;,\quad \deg\Phi_a=0\;,  
\end{equation}  
and enumerated by Latin characters $a,b,c,\ldots$. The other one consists   
of odd constraints, for which we will use the notation  
\begin{equation}  
\chi_\alpha\equiv\chi_\alpha(f^{\dagger},f)=  
\,\,:\!\chi_\alpha(f^{\dagger},f)\!:\,\,=\chi_\alpha^{\dagger}\;,  
\quad \deg\chi_\alpha=1\;.  
\end{equation}  
They will be enumerated by Greek letters $\alpha ,\beta ,\gamma   
,\ldots$. With these constraints the physical Hilbert space is determined  
by the conditions  
\begin{equation}  
\Phi_a|\varphi\rangle_{\rm phys}=0\;,
\quad\chi_\alpha|\varphi\rangle_{\rm phys}=0\;, 
\label{23} 
\end{equation}  
for all $a$ and $\alpha $. Note that here we have assumed that the   
constraint operators are self-adjoint. If they are not self-adjoint we   
will assume that they appear in pairs such as $(\chi ,\chi^{\dagger})$  
which in turn allows us to generate self-adjoint constraints   
via proper linear combinations like $\chi +\chi ^{\dagger}$ and   
$\I\chi -\I\chi ^{\dagger}$.   
  
Following Dirac \cite{Dirac64} we group the constraints into two classes.  
For first-class constraints the above conditions (\ref{23}) need to be  
enforced only  
initially at $t=0$ as the quantum evolution guarantees that a physical  
state will always remain in the physical Hilbert space as time evolves.  
If this is not the case there exists at least one constraint which is of  
second class.  
  
The above characterization of first-class constraints is equivalent to  
the requirement that they obey the following commutation and    
anticommutation relations.  
\begin{equation}
\begin{array}{l}  
[\Phi_a,\Phi_b]:=\Phi_a\Phi_b-\Phi_b\Phi_a=\I c_{ab}{}^c\Phi_c\;,\\[2mm]
[\Phi_a,\chi_\alpha]=\I d_{a\alpha}{}^\beta\chi_\beta\;,\quad
\{\chi_\alpha,\chi_\beta\}=\I g_{\alpha\beta}{}^a\Phi_a\;. 
\label{24}
\end{array}
\end{equation}
\begin{equation}  
[\Phi_a,H]=\I h_{a}{}^b\Phi_b\;,\quad   
[\chi_\alpha,H]=\I k_{\alpha}{}^\beta\chi_\beta\;.  
\label{25}
\end{equation}  
In other words, the constraints together with the Hamiltonian form a Lie   
superalgebra \cite{Corn89} defined by the structure constants $c,d,g,h$   
and $k$. 
In general these structure constants could be operator-valued quantities
depending on the fermion operators. Throughout this paper we will, however,
consider only thoses cases where the structure constants are complex valued 
numbers. 
Let us also note that the first-class constraints alone define a Lie   
superalgebra (\ref{24}) which is an ideal of the total algebra including 
(\ref{25}). 
This ideal generates a Lie supergroup (via the usual exponential map) 
which in turn   
would enable us to construct in combination with the associated   
invariant Haar measure \cite{WilCor84} a proper projection operator in   
analogy to the approach of \cite{Klauder97}. However, things are much  
simpler in this case. In particular, with the help of   
the last anticommutation relation in (\ref{24}) one can easily show that the   
first condition in (\ref{23}), that is, $\Phi _{a}|\psi \rangle_{\rm phys}=0$
for all $a$, implies the second one. In other words, in the case of   
first-class constraints the odd constraints are implied by the even   
constraints. This argument holds only for the case when even constraints   
are present. If this would not be the case, then the algebra of the   
constraints  reduces to $\{\chi_{\alpha},\chi _{\beta }\}=0$ for all $\alpha$ 
and $\beta$. This algebra, however, does not have a non-trivial 
(in)finite-dimensional realization. Actually, 
such an algebra implies $\chi_\alpha|\psi\rangle=0$ 
for all $\alpha$ and all $\psi\in{\cal H}$. 
Or in other words, the only possible self-adjoint realization of purely odd 
first-class constraints are given by $\chi_\alpha \equiv 0$, and hence does 
not represent any constraints.   
  
\subsection{The projection operator}  
Because of the above mentioned properties it suffices to consider only   
the ordinary Lie algebra spanned by the even constraints $\{\Phi_a\}$ 
with structure constants $c_{ab}{}^c$. 
We may construct a proper projection operator via 
the invariant Haar measure of the   
corresponding Lie group following \cite{Klauder97}. Let us be more   
explicit. The general group element generated by the even constraints is   
given by   
\begin{equation}  
\exp\{-\I \xi^a\Phi_a(f^{\dagger},f)\}\;,  
\label{26}
\end{equation}  
where $\{\xi^a\}$ are real group parameters. To be more precise, 
(\ref{26}) is a $2^N$-dimensional unitary fully reducible representation of
this Lie group in ${\cal H}$. 
For simplicity, we consider here only the case of a compact group. For the
treatment in cases of non-compact groups see \cite{Comment}.
For a compact group,
let us denote the corresponding invariant normalized Haar measure by   
$d\mu(\xi)$. Then a proper projection operator may be defined by 
\cite{BarRac80}
\begin{equation}  
\e:=\int d\mu(\xi)\,  
\exp\{-\I \xi^a\Phi_a\}  
\end{equation}  
which due to the invariance of the Haar measure and the group-composition law  
obviously obeys the properties ${\e}={\e}^2={\e}^{\dagger}$ of an orthogonal   
projector. It projects onto the physical Hilbert space since by construction 
the
physical states are the eigenstates of ${\e}$ with eigenvalue one,  
${\e}|\psi\rangle_{\rm phys}=|\psi\rangle_{\rm phys}$. Furthermore, we   
note that  
\begin{equation}  
\exp\{-\I \xi^a\Phi_a\}{\e}={\e}  
\end{equation}  
for any set $\{\xi^a\}$ and   
\begin{equation}  
\E^{-\I tH}{\e}={\e}\E^{-\I tH}={\e}\E^{-\I tH}{\e}=  
{\e}\E^{-\I t({\e}H{\e})}{\e}\;,  
\label{29} 
\end{equation}  
which is the (constrained) time-evolution operator in the physical subspace.  
As an aside we mention that this operator may be viewed as an element of the  
Lie group, associated with the Lie algebra spanned by the Hamiltonian   
and the even constraints, which is   
averaged over the subgroup associated with the subalgebra of the even   
constraints. In other words, it is invariant under right and left   
multiplication of this subgroup and, hence, belongs to the corresponding   
two-sided coset.  
 
Finally, let us mention that the $N$-fermion Hilbert space is finite 
dimensional and, hence, the spectrum of the constraints is pure point. 
Therefore, technical difficulties arising from a possible continuous spectrum  
of the constraints (see ref.\ \cite{Klauder97}) do not occur. 
 
\subsection{\protect Path-integral representations for the constrained
propagator}
Let us now construct a path-integral representation for the  
constrained propagator, that is, the coherent-state matrix element of the   
constrained time-evolution operator (\ref{29}):  
\begin{equation}  
\begin{array}{l}  
\langle\psi''|\E^{-\I tH}{\e}|\psi'\rangle  =   
\langle\psi''|\E^{-\I tH}\E^{-\I \xi^a\Phi_a}  
{\e}|\psi'\rangle\\[2mm]  
\quad = \displaystyle \int d\bar{\psi }_{0}d\psi _{0}\,   
\langle\psi''|\E^{-\I tH}\E^{-\I \xi^a\Phi_a}|\psi_{0}\rangle   
\langle\psi_{0}|{\e}|\psi'\rangle \;. 
\end{array}  
\end{equation}  
Making use of the group composition law, which follows from the algebra   
of the even constraints, setting again $\varepsilon=t/N$ and inserting the 
resolution  (\ref{12}) of the identity we find  
\begin{equation}  
\begin{array}{l}   
\displaystyle   
\langle\psi''|\E^{-\I tH}\E^{-\I \xi^a\Phi_a}|\psi _{0}\rangle\\[2mm]
\quad=\displaystyle  
\langle\psi _{N}|\prod_{n=1}^{\mbox{\small $N$} \atop \longleftarrow}
\left(\E^{-\I \varepsilon H}
\E^{-\I \varepsilon\eta_n^a\Phi_a}\right)|\psi _{0}\rangle\\[2mm]   
\quad\displaystyle   
=\prod_{n=1}^{N-1}\int d\bar{\psi }_{n}d\psi _{n}   
\prod_{n=1}^{N}\langle \psi _{n}|\E^{-\I \varepsilon H}   
\E^{-\I \varepsilon\eta_n^a\Phi_a}|\psi _{n-1}\rangle\;,  
\end{array}   
\end{equation}  
where $\{\eta_n^a\}$ are appropriate real numbers.   
Taking, as in Sect.\ 2.3, the limit $\varepsilon\to 0$ one ends up with  
the following time-lattice definition of a constrained fermion coherent-state  
path integral (notation as in Sect.\ 2.3 except $\psi'\neq\psi_0$)  
\begin{equation}  
\begin{array}{l}   
\displaystyle   
\langle \psi''|\E^{-\I tH}{\e}|\psi'\rangle=\displaystyle   
\lim_{\varepsilon\to0}  
\prod_{n=0}^{N-1}\int d\bar{\psi }_{n}d\psi _{n}\int d\mu(\xi)\\[2mm]  
\quad\displaystyle\times   
\exp\left\{-\sum_{n=1}^{N}\left[  
\frac{1}{2}\,\bar{\psi}_{n}\Delta\psi_{n}-\frac{1}{2}\,
\Delta\bar{\psi}_{n}\psi_{n-1}\right.\right.\\  
\qquad\left.\rule{0mm}{6mm}\left.\rule{0mm}{5mm}
+\I\varepsilon  H(\bar{\psi }_{n},\psi _{n-1})  
+\I\varepsilon\eta^a_{n}\Phi_a(\bar{\psi }_{n},\psi _{n-1})
\right]\right\}\\[2mm]  
\quad\displaystyle   
\times\langle   
\psi_{0}|\exp\{-\I \xi^a\Phi_a(f^{\dagger},f)\}  
|\psi'\rangle   \;.
\end{array} \label{32}  
\end{equation}  
Hence, we arrive at the formal path-integral representation of the constrained
propagator  
\begin{equation}  
\begin{array}{l}  
\displaystyle  
\langle \psi''|\E^{-\I tH}{\e}|\psi'\rangle
=\int{\cal D}\bar{\psi }{\cal D}\psi\int d\mu(\xi)\\[2mm]
\quad\times\displaystyle  
\exp\left\{\I\int_{0}^t d\tau \left[\frac{\I}{2}(\bar{\psi }\dot{\psi }-   
\dot{\bar{\psi }}\psi )-H(\bar{\psi },\psi )\right.\right.\\[2mm]
\qquad\displaystyle
\left.\rule{0mm}{5mm}\left.\rule{0mm}{5mm}
-\eta^a\Phi_a(\bar{\psi },\psi )\right]\right\}
\exp\left\{-\I \xi^a\Phi_a(\bar{\psi}',\psi')\right\}\;.
\end{array} \label{33}  
\end{equation}  
Despite the fact that in this path integral  the   
time-dependent real-valued functions $\{\eta^a\}$ explicitly appear, which 
may be interpreted as   
Lagrange multipliers, it is completely independent of them as is clearly  
shown by the left-hand side. Hence, as in \cite{Klauder97}, we are free to  
average the right-hand side over the functions $\{\eta^a\}$  
with an arbitrary in general complex-valued measure $C(\eta)$ 
which is normalized, $\int{\cal D}C(\eta)=1$. 
The only requirement we impose on this measure is,
that such an average will introduce at least one projection operator   
${\e}$ to account for the initial value equation (\ref{23}). If it puts in two
or more of these projection operators the result will be the same since 
$\e^2=\e$.
Hence, there are many forms for this measure which will be admissible. For an
example see the Appendix.
In doing so we have derived yet another path-integral   
representation of the constrained propagator.  
\begin{equation}
\begin{array}{l}  
\displaystyle\langle \psi''|\E^{-\I tH}{\e}|\psi'\rangle=  
\int{\cal D}\bar{\psi }{\cal D}\psi\!\int{\cal D}C(\eta)
\exp\left\{\I\int_{0}^t d\tau\right.\\[3mm]  
\quad\displaystyle\times
\left.\left[\frac{\I}{2}(\bar{\psi }\dot{\psi }-   
\dot{\bar{\psi }}\psi )-H(\bar{\psi },\psi )-   
\eta^a\Phi_a (\bar{\psi },\psi )\right]\right\}\;.  
\label{34} 
\end{array}
\end{equation}  
In essence, formulas (\ref{32}), (\ref{33}) and (\ref{34}) resemble the 
fermionic counter parts of the results (64), (65) and (66) in \cite{Klauder97}
where the bosonic case has been studied. 

\section{Examples of first-class constraints}  
As we have seen in the above discussion, the treatment of first-class   
constraints for fermionic systems is very much the same as that for   
bosonic systems \cite{Klauder97}. In particular, it is sufficient to   
consider only even constraints which are bosonic in nature. Therefore, we   
will discuss below only two examples which demonstrate the minor   
differences to the bosonic case.  
  
\subsection{First example of first-class constraints}  
As a simple example with purely even constraints let us consider a   
system of an $N$ fermion system subjected to the even constraint   
\begin{equation}  
\Phi(f^{\dagger},f)=\sum_{i=1}^Nf_i^{\dagger}f_i-M\;.  
\end{equation}  
Obviously, this constraint fixes the number of fermions to $M\in   
{\Bbb N}$ with $M\leq N$. In order to make the effects of the   
constraints more transparent we will consider only the   
path-integral representation of the coherent-state matrix element of the   
projection operator   
\begin{equation}  
{\e}=\int_0^{2\pi}\frac{d\xi}{2\pi}\,\E^{-\I \xi\Phi}=\delta _{\Phi ,0}
={\e}^2={\e}^{\dagger}\;,  
\label{36} 
\end{equation}  
that is, we will consider a system with a vanishing Hamiltonian, $H=0$, and 
limit ourselves to the special case $M=1$, $N=2$.  
Formally, the corresponding path integral is then given by  
\begin{equation}  
\begin{array}{l}
\displaystyle
\int{\cal D}\bar{\Psi }{\cal D}\Psi \int{\cal D}C(\eta )\\[2mm]
\quad\displaystyle\times
\exp\left\{\I\int_{0}^{t}d\tau   
\left[\frac{\I}{2}(\bar{\Psi }\cdot\dot{\Psi }-\dot{\bar{\Psi }}\cdot\Psi)  
-\eta (\bar{\Psi }\cdot\Psi -1)\right]\right\}  
\end{array}
\label{38}
\end{equation}  
and leads to the coherent-state matrix element (for details see the Appendix)
\begin{equation}  
\langle\Psi ''|{\e}|\Psi '\rangle=\E^{-\bar{\Psi }''\cdot\Psi ''/2}\,  
\E^{-\bar{\Psi }'\cdot\Psi '/2}\,\bar{\Psi }''\cdot\Psi '  
\label{39}
\end{equation}  
where we have adopted the short-hand notation of (\ref{17}).  
We leave it to the reader to verify that this   
matrix element represents a reproducing   
kernel in the physical subspace given by the   
linear span of the two vectors $|01\rangle$ and $|10\rangle$:  
\begin{equation}  
\int d\bar{\Psi }d\Psi \,\langle\Psi ''|{\e}|\Psi \rangle  
\langle\Psi |{\e}|\Psi' \rangle=  
\langle\Psi ''|{\e}|\Psi' \rangle\;,  
\end{equation}  
where   
$d\bar{\Psi }d\Psi:=d\bar{\psi}_{1}d\psi_{1}d\bar{\psi}_{2}d\psi_{2}$.  
  
\subsection{Second example of first-class constraints}  
As a second example we will now consider a three-fermion system ($N=3$) 
subjected to one even and two odd constraints given by  
\begin{equation}  
\begin{array}{l}  
\Phi =1-f^{\dagger}_{1}f_{1}-f^{\dagger}_{2}f_{2}  
       -f^{\dagger}_{2}f_{2}+f^{\dagger}_{1}f_{1}f^{\dagger}_{2}f_{2}\\[2mm]
\qquad +f^{\dagger}_{2}f_{2}f^{\dagger}_{3}f_{3}  
       +f^{\dagger}_{3}f_{3}f^{\dagger}_{1}f_{1}\;,\\[2mm]  
\chi=f_1f_2f_3\;,\quad   
\chi^{\dagger}=f^{\dagger}_3f^{\dagger}_2f^{\dagger}_1\;.  
\end{array}  
\end{equation}  
These first-class constraints obey the Lie superalgebra  
\begin{equation}  
[\chi,\Phi]=0=[\chi^{\dagger},\Phi]\;,\quad  
\{\chi,\chi^{\dagger}\}=\Phi\;,\quad   
\chi^2=0=(\chi^{\dagger})^2.  
\end{equation}  
Obviously, the six-dimensional physical subspace is characterized by   
having at least one empty and one occupied fermion state.  
As in the previous example the spectrum of the even constraint $\Phi $   
is integer and therefore the projection operator has the same integral   
representation.  
\begin{equation}  
{\e}=\int_0^{2\pi}\frac{d\xi}{2\pi}\,\E^{-\I \xi\Phi}  
\end{equation}  
and can explicitly be expressed in terms of the fermion number operators  
\begin{equation}
\begin{array}{ll}
{\e}&=f^{\dagger}_{1}f_{1}(1-f^{\dagger}_{2}f_{2})+  
f^{\dagger}_{2}f_{2}(1-f^{\dagger}_{3}f_{3})+  
f^{\dagger}_{3}f_{3}(1-f^{\dagger}_{1}f_{1})\\[2mm]
&=1-\Phi \;.  
\end{array}
\end{equation}  
The path integral for the coherent-state matrix element of the projection 
operator formally reads 
\begin{equation} 
\langle\Psi ''|{\e}|\Psi '\rangle=\int{\cal D}\bar{\Psi}{\cal D}\Psi 
\int{\cal D}C(\eta)\exp\left\{\I\int_0^t d\tau\,L\right\}\;, 
\end{equation} 
where 
\begin{equation}
\begin{array}{l} 
L:=\textstyle\frac{\I}{2} 
\Bigl(\bar{\Psi}\cdot\dot{\Psi} -\dot{\bar{\Psi}}\cdot\Psi\Bigr)
-\eta\Bigl(1-\bar{\Psi}\cdot\Psi\\[2mm]
\qquad+\bar{\psi}_1\psi_1\bar{\psi}_2\psi_2 
+\bar{\psi}_2\psi_2\bar{\psi}_3\psi_3+\bar{\psi}_3\psi_3\bar{\psi}_1\psi_1 
\Bigr)\;. 
\end{array}
\end{equation} 
An explicit path integration then leads to the result  
\begin{equation}  
\begin{array}{l}  
\langle\Psi ''|{\e}|\Psi '\rangle=\langle\Psi ''|\Psi '\rangle
\Bigl[\bar{\Psi}''\cdot\Psi'- 
\bar{\psi }''_{1}\psi '_{1}\bar{\psi }''_{2}\psi '_{2}\\[2mm]  
\qquad  
-\bar{\psi }''_{2}\psi '_{2}\bar{\psi }''_{3}\psi '_{3}- 
\bar{\psi }''_{3}\psi '_{3}\bar{\psi }''_{1}\psi '_{1} 
\Bigr]\\[2mm]  
\quad=\E^{-(\bar{\Psi}''\cdot\Psi''+\bar{\Psi}'\cdot\Psi')/2}
\Bigl[\bar{\Psi}''\cdot\Psi'+ 
\bar{\psi }''_{1}\psi '_{1}\bar{\psi }''_{2}\psi '_{2}\\[2mm]
\qquad+ 
\bar{\psi }''_{2}\psi '_{2}\bar{\psi }''_{3}\psi '_{3}+ 
\bar{\psi }''_{3}\psi '_{3}\bar{\psi }''_{1}\psi '_{1} 
\Bigr] \;. 
\end{array} 
\end{equation}  
 
\section{Second-class constraints}  
Second-class constraints are all those which are not first class. For   
second-class constraints it is not sufficient to start with an initial   
state on the physical subspace as in this case the time evolution   
generated by the Hamiltonian will generally depart from the physical 
subspace. In   
other words, after some short time interval (say $\varepsilon$)  
one may have to project the state back onto the physical subspace. Hence, we   
are led to consider the constrained propagator  
\begin{equation}  
\begin{array}{l}  
\displaystyle  
\langle\psi''|{\e}\E^{-\I t({\e}H{\e})}  
{\e}|\psi'\rangle \\[2mm]
\quad=\displaystyle  
\lim_{\varepsilon\to 0}\langle\psi''|{\e}\E^{-\I \varepsilon H}  
{\e}\E^{-\I \varepsilon H}{\e}\cdots {\e}\E^{-\I \varepsilon H}{\e}  
|\psi'\rangle \\[2mm]  
\quad\displaystyle=\lim_{\varepsilon \to 0}  
\int\prod_{n=1}^{N-1}d\bar{\psi }_{n}d\psi _{n}  
\prod_{n=1}^{N}\langle\psi_{n}|{\e}\E^{-\I \varepsilon H}{\e}  
|\psi_{n-1}\rangle\;.   
\end{array} 
\label{45}  
\end{equation}  
Again we will closely follow the basic ideas used in the canonical   
coherent-state path-integral approach \cite{Klauder97}. Hence, we
start by introducing the unit vectors   
$|\psi\rangle\rangle:={\e}|\psi\rangle/||{\e}|\psi\rangle||$  
and set $M'':=||{\e}|\psi'' \rangle ||$,  
$M':=||{\e}|\psi'\rangle ||$. The path integral for the constrained propagator
can then be rewritten as  
\begin{equation}  
\begin{array}{l}
\displaystyle
M''M'\lim_{\varepsilon \to 0}  
\int\left[\prod_{n=1}^{N-1}d\bar{\psi }_{n}d\psi _{n}\,
\langle\psi_{n}|{\e}|\psi_{n}\rangle\right]  \\[2mm]
\quad\displaystyle\times
\prod_{n=1}^{N}  
\langle\langle\psi_{n}|\E^{-\I \varepsilon H}|\psi_{n-1}\rangle\rangle  
\label{46} 
\end{array}
\end{equation}  
which admits the following formal path-integral representation  
\begin{equation}
\begin{array}{l}  
\displaystyle
M''M'\int{\cal D}_{E}\mu(\bar{\psi },\psi )\\[2mm]
\quad\displaystyle\times  
\exp\left\{\I\int_{0}^{t}d\tau   
\Bigl[\I\langle\langle\psi |\textstyle\frac{d}{d\tau }|\psi \rangle\rangle-  
\langle\langle\psi |H|\psi \rangle\rangle\Bigr]\right\}\;.  
\label{47} 
\end{array}
\end{equation}  
In terms of the original vectors it reads  
\begin{equation}  
\begin{array}{l}
\displaystyle
M''M'\int{\cal D}_{E}\mu(\bar{\psi },\psi )\\[2mm]  
\quad\displaystyle\times
\exp\left\{\I\int_{0}^{t}d\tau   
\left[\I\frac{\langle\psi |\frac{d}{d\tau }|\psi\rangle}  
{\langle\psi |{\e}|\psi\rangle}-  
\frac{\langle\psi |H|\psi\rangle}{\langle\psi |{\e}|\psi\rangle}  
\right]\right\}\;. 
\label{48} 
\end{array}
\end{equation}  
  
Another relation may be obtained by assuming that the projection   
operator allows for an integral representation in terms of the even and   
odd constraints  
\begin{equation}  
{\e}=\int d\mu_{\varepsilon }(\eta ,\lambda )\,  
\E^{-\I \varepsilon (\eta ^{a}\Phi_{a}+\lambda ^{\alpha }\chi_{\alpha })}  
\label{49} 
\end{equation}  
where $d\mu_\varepsilon $ stands for some even Grassmann-valued  
measure depending on the real variables $\eta ^{\alpha }$  
and the odd Grassmann numbers $\lambda  ^{\alpha }$  
which both may be considered as Lagrange multipliers.   
Using this relation in the path-integral expression (\ref{45}) we   
find the representation (notation as in Sect.\ 2.3 except 
$\psi _{N}\neq \psi ''$)  
\begin{equation}  
\begin{array}{l}  
\displaystyle  
\lim_{\varepsilon \to 0}  
\int\left[\prod_{n=1}^{N}  
d\bar{\psi }_{n}d\psi _{n}d\mu _{\varepsilon }(\eta _{n},\lambda _{n})\right]  
d\mu _{\varepsilon }(\eta _{0},\lambda _{0})\\[4mm]  
\quad\displaystyle  
\times\langle\psi''|  
\E^{-\I \varepsilon(\eta_{N}^{a}\Phi_{a}+\lambda_{N}^{\alpha}\chi_{\alpha })}  
|\psi_{N}\rangle\\[2mm]
\quad\displaystyle\times
\prod_{n=1}^{N}\langle\psi_{n}|\E^{-\I \varepsilon H}  
\E^{-\I \varepsilon(\eta_{n-1}^{a}\Phi_{a}
+\lambda_{n-1}^{\alpha}\chi_{\alpha})}|\psi_{n-1}\rangle  
\end{array}  
\end{equation}  
which can formally be written as  
\begin{equation}
\begin{array}{l}
\displaystyle
\int{\cal D}\bar{\psi }{\cal D}\psi {\cal D}E(\eta ,\lambda )  
\exp\left\{\I\int_{0}^{t}d\tau   
\left[\frac{\I}{2}(\bar{\psi }\dot{\psi }-\dot{\bar{\psi}}\psi )
\right.\right.\\[2mm]
\displaystyle\left.\rule{0mm}{6mm}\left.\rule{0mm}{5mm}
\,\,- H(\bar{\psi },\psi )-\eta ^{a}\Phi_{a}(\bar{\psi },\psi )-  
\lambda ^{\alpha }\chi _{\alpha }(\bar{\psi },\psi )\right]\right\}\;.  
\label{51} 
\end{array}
\end{equation}  
Here let us remark that we have assumed that the constraints are  
self-adjoint. This is typically not the case for odd constraints, which   
then appear in pairs $(\chi,\chi ^{\dagger})$. As a consequence the   
Grassmann-valued Lagrange multipliers also appear in pairs   
$(\lambda,\bar{\lambda })$. In contrast to the first-class   
constraints, in the present case one cannot neglect the odd constraints.   
However, the appearance of Grassmann multipliers may be omitted at the   
expense of no longer having the constraints appear explicitly in the   
exponent of (\ref{49}). Actually, because ${\rm spec}({\e})\subseteq\{0,1\}$
we may always choose the following simple integral representation   
of the projection operator  
\begin{equation}  
{\e}=\int_{0}^{2\pi }\frac{d\eta }{2\pi}\,  
\E^{-\I \eta (1-{\e})}\;.  
\label{52}
\end{equation}  
 
Again we would like to point out that eqs.\ (\ref{45})-(\ref{48}) are the  
fermion counterparts of eqs.\ (104)-(106) of ref.\ \cite{Klauder97}, and 
relation (\ref{51}) corresponds to (109) in \cite{Klauder97}. 
 
\section{Examples of second-class constraints}  
Even fermionic constraints are in essence similar to bosonic constraints   
which have extensively been discussed in \cite{Klauder97}. For this   
reason we will concentrate our attention in this section exclusively on   
odd second-class constraints. We will start with two simple examples of   
constraints linear in fermion operators and then generalize our   
approach to an arbitrary set of linear constraints. Based on an example   
of a non-linear odd constraint we will show that all non-linear 
diagonal odd constraints can be reduced to the linear case.  
 
\subsection{Linear odd constraints}  
As mentioned above we will begin our discussion with a simple, that is $N=1$,
fermion system which obeys the constraints  
\begin{equation}  
\chi=f-\theta\;,\quad \chi^{\dagger}=f^{\dagger}-\bar{\theta }\;.
\label{53} 
\end{equation}  
Here $\bar{\theta},\theta\in{\Bbb C}B_2$ are odd 
Grassmann numbers. The constraints (\ref{53}) obey the following 
anticommutation relations 
\begin{equation}  
\{\chi ,\chi ^{\dagger}\}=1\;,\quad \chi ^{2}=0=(\chi ^{\dagger})^{2}  
\label{54} 
\end{equation}  
and, therefore, one cannot impose both constraint conditions  
\begin{equation}  
\begin{array}{ll}  
\chi|\varphi \rangle_{\rm phys}=0\;,\qquad & \mbox{case A}\\[2mm]  
\chi^{\dagger}|\varphi \rangle_{\rm phys}=0\;,& \mbox{case B}  
\end{array}  
\label{55} 
\end{equation}  
simultaneously. Such a procedure would clearly lead to an inconsistent quantum
theory. There are several ways to relax the conditions in order to
formulate a consistent approach.
Here we adopt an approach similar to the so-called holomorphic
quantization \cite{HenTei92} utilized for bosonic models with similar
constraint inconsistencies. That is, we will consider only one of the above 
two conditions to define a proper physical Hilbert subspace. 
However, both possible cases will be discussed for completeness.
  
\subsubsection{Case A}
The solution of (\ref{55}) in case A is obviously given by the   
fermion coherent state $|\theta \rangle$ and the corresponding projection   
operator reads  
\begin{equation}  
\begin{array}{ll}
\e_{A}&\displaystyle
=|\theta \rangle\langle\theta |=\chi \chi ^{\dagger}=  
\int d\bar{\lambda}d\lambda\,   
\E^{-\I \bar{\lambda}\chi}\E^{-\I \chi^{\dagger}\lambda} \\[2mm]
&\displaystyle
=\int d\bar{\lambda}d\lambda\,   
\E^{\bar{\lambda}\lambda/2}\,\E^{-\I (\bar{\lambda}\chi+
\chi^{\dagger}\lambda)}\;.  
\label{56} 
\end{array}
\end{equation}  
The diagonal coherent-state matrix element of this operator, needed for   
example in evaluating the path integral (\ref{46}), is given by  
\begin{equation}  
\langle\psi _{n}|{\e }_{A}|\psi _{n}\rangle=  
\exp\{-(\bar{\psi } _{n}-\bar{\theta })(\psi _{n}-\theta )\}\;.  
\end{equation}  
Hence, for a normal-ordered Hamiltonian $H=H(f^{\dagger},f)$ we arrive at the
formal path-integral expressions for the constrained propagator  
\begin{equation}  
\begin{array}{l}  
\displaystyle  
\int{\cal D}\bar{\psi }{\cal D}\psi {\cal D}E(\bar{\lambda },\lambda )  
\exp\left\{\I\int_{0}^{t}d\tau   
\left[\frac{\I}{2}(\bar{\psi }\dot{\psi }-\dot{\bar{\psi}}\psi )
\right.\right.\\[2mm]
\quad\displaystyle\left.\rule{0mm}{5mm}\left.\rule{0mm}{5mm}
-H(\bar{\psi },\psi )-\bar{\lambda}(\psi-\theta )  
-(\bar{\psi}-\bar{\theta })\lambda \right]\right\}\\[2mm]  
\quad\displaystyle=  
\int{\cal D}\bar{\psi }{\cal D}\psi   
\exp\left\{\I\int_{0}^{t}d\tau   
\left[\frac{\I}{2}(\bar{\psi }\dot{\psi }-\dot{\bar{\psi}}\psi )
\right.\right.\\[2mm]
\quad\displaystyle\left.\rule{0mm}{5mm}\left.\rule{0mm}{5mm}  
+\I(\bar{\psi }-\bar{\theta })(\psi -\theta )-  
H(\bar{\psi },\psi )\right]\right\}\;.  
\end{array}  
\label{57}
\end{equation}  
Explicit path integration (see Appendix) will then lead to the final result  
\begin{equation}  
\begin{array}{l}  
\displaystyle  
\langle\psi ''|{\e}_{A}\E^{-\I t({\e}_{A}H{\e}_{A})}{\e}_{A}  
|\psi '\rangle  
=\displaystyle  
\langle\psi ''|\theta\rangle\langle\theta|\psi'\rangle  
\E^{-\I tH(\bar{\theta },\theta )}\\[2mm]  
\quad=\displaystyle  
\langle\psi ''|\psi'\rangle  
\exp\left\{-(\bar{\psi}''-\bar{\theta })(\psi '-\theta )  
-\I tH(\bar{\theta },\theta )\right\}\;.  
\end{array}  
\label{58}
\end{equation}  

\subsubsection{Case B}  
For the second choice (case B) the solution of (\ref{55}) is given by a   
different kind of coherent states defined by \cite{OhnKas78,EzaKla85}  
\begin{equation}  
|\varphi\rangle_{\rm phys}=|\bar{\theta}):=\E^{\bar{\theta}\theta/2}  
\Bigl(|1\rangle-\bar{\theta }|0\rangle\Bigr)\;.  
\end{equation}  
In contrast to the even fermion coherent states introduced in Sect.\ 2.2,   
these states are odd. They are eigenstates of the fermion   
creation operator and are orthogonal to the corresponding even states:  
\begin{equation}  
f^{\dagger}|\bar{\theta })=\bar{\theta }|\bar{\theta })\;,\quad  
(\bar{\theta}|f=(\bar{\theta}|\theta\;,\quad\langle\theta|\bar{\theta})=0\;.  
\end{equation}  
For case B the projection operator is given by the orthogonal complement of   
(\ref{56})  
\begin{equation}
\begin{array}{ll}  
\e_{B}&\displaystyle
=|\bar{\theta })(\bar{\theta }|=\chi ^{\dagger}\chi=  
\int d\bar{\lambda}d\lambda\,   
\E^{\I\chi^{\dagger}\lambda}\E^{\I\bar{\lambda}\chi}\\[2mm]  
&\displaystyle
=\int d\bar{\lambda}d\lambda\,   
\E^{-\bar{\lambda}\lambda/2}\,\E^{\I(\bar{\lambda}\chi
+\chi^{\dagger}\lambda)}={\bf 1}-\e_A 
\label{62} 
\end{array}
\end{equation}  
whose diagonal coherent-state matrix element reads  
\begin{equation}  
\langle\psi _{n}|{\e }_{B}|\psi _{n}\rangle=  
(\bar{\psi }_{n}-\bar{\theta })(\psi _{n}-\theta )\;.  
\end{equation}  
Explicit path integration will then lead to the constrained propagator   
\begin{equation}  
\begin{array}{l} 
\langle\psi ''|{\e}_{B}\E^{-\I t({\e}_{B}H{\e}_{B})}  
{\e}_{B}|\psi '\rangle =  
\langle\psi ''|\bar{\theta })(\bar{\theta }|\psi'\rangle  
\E^{-\I th(\theta ,\bar{\theta })}\\[2mm] 
\quad=\langle\psi ''|\psi'\rangle (\bar{\psi}''-\bar{\theta })(\psi'-\theta ) 
\E^{-\I th(\theta ,\bar{\theta })}\;,  
\end{array} 
\label{63}
\end{equation}  
where $h(\theta ,\bar{\theta }):=(\bar{\theta }|H|\bar{\theta })$. Note   
that for an {\em anti}-normal ordered Hamiltonian $H=H(f,f^{\dagger})$ we have
$h(\theta ,\bar{\theta })=H(\theta ,\bar{\theta })$.
\pagebreak[1]  

\subsubsection{A second example}  
As a second example of linear constraints let us consider an $N=2$   
fermion system subjected to the two odd constraints  
\begin{equation}  
\chi =\frac{1}{\sqrt{2}}(f_{1}-f_{2})\;,\quad   
\chi^{\dagger} =\frac{1}{\sqrt{2}}(f^{\dagger}_{1}-f^{\dagger}_{2})\;,  
\end{equation}  
which also obey the algebra (\ref{54}). In analogy to the previous example  
we may again consider two different physical subspaces according to   
case A and B in (\ref{55}).  
  
For case A the physical Hilbert space is the two-dimensional subspace   
spanned by the fermion number eigenstates    
$|00\rangle$ and $(|01\rangle+|10\rangle)/\sqrt{2}$. The   
corresponding projection operator is given by $\e_{A}=\chi \chi   
^{\dagger}$ and admits integral representations as given in (\ref{56}). The   
path integral for its matrix element (for simplicity we consider here the   
system $H=0$) leads to  
\begin{equation}  
\begin{array}{l}  
\langle\psi _{1}''\psi _{2}''|\e_{A}|\psi _{1}'\psi _{2}'\rangle\\[2mm]
\quad=\langle\psi _{1}''\psi _{2}''|\psi _{1}'\psi _{2}'\rangle  
\left[1-\frac{1}{2}(\bar{\psi }_{1}''-\bar{\psi }_{2}'')  
(\psi _{1}'-\psi _{2}')\right]\\[2mm]  
\quad=\E^{-\bar{\Psi}''\cdot\Psi''/2}\,\E^{-\bar{\Psi}'\cdot\Psi'/2}  
\left[1+\frac{1}{2}(\bar{\psi }_{1}''+\bar{\psi }_{2}'')  
(\psi _{1}'+\psi _{2}')\right].  
\end{array}  
\label{65}
\end{equation}  
  
In case B we are dealing with the projection operator $\e_{B}={\bf 1}-  
\e_{A}=\chi ^{\dagger}\chi $ and its integral representations are the same   
as in (\ref{62}). This operator projects onto the orthogonal complement of the
previous case, that is, onto the subspace spanned by  
$|11\rangle$ and $(|01\rangle-|10\rangle)/\sqrt{2}$.   
Here the result of path integration for the coherent-state matrix   
element of $\e_{B}$ reads  
\begin{equation}  
\begin{array}{l}  
\langle\psi _{1}''\psi _{2}''|\e_{B}|\psi _{1}'\psi _{2}'\rangle=  
\langle\psi _{1}''\psi _{2}''|\psi _{1}'\psi _{2}'\rangle  
\frac{1}{2}(\bar{\psi}_{1}''-\bar{\psi}_{2}'')(\psi_{1}'-\psi_{2}')\\[2mm]  
\quad=\E^{-(\bar{\Psi}''\cdot\Psi''+\bar{\Psi}'\cdot\Psi')/2}\\[2mm]
\qquad\times
\left[\bar{\psi }''_{1}\psi '_{1}\bar{\psi }''_{2}\psi '_{2}+  
\frac{1}{2}(\bar{\psi }_{1}''-\bar{\psi }_{2}'')  
(\psi _{1}'-\psi _{2}')\right].  
\end{array}
\label{66}  
\end{equation}  

\subsubsection{Generalization}  
The above discussion may easily be generalized to a set of diagonal linear   
second-class constraints obeying the anticommutation relations  
\begin{equation}  
\{\chi_\alpha,\chi_\beta\}=0=\{\chi_\alpha^{\dagger},\chi_\beta^{\dagger}\}\;,
\quad\{\chi_\alpha,\chi_\beta^{\dagger}\}=\delta_{\alpha\beta}\;,  
\label{68}
\end{equation}  
where $\alpha ,\beta \in\{1,2,\ldots,M \}$, $M\leq N$.  
Clearly, for each $\alpha $ one has two choices for a projection operator,   
$\e_{A}^{(\alpha )}=\chi _{\alpha }\chi^{\dagger} _{\alpha }$ or
$\e_{B}^{(\alpha )}=\chi^{\dagger} _{\alpha }\chi_{\alpha }$. Therefore,   
for the total physical subspace the corresponding projection operator is   
not unique and we have to choose one out of the following $2^{M}$   
possible operators,  
\begin{equation}  
\e=\e^{(1)}_{i_{1}}\e^{(2)}_{i_{2}}\cdots\e^{(M)}_{i_{M}}\;,\quad  
i_{\alpha }\in \{A,B\}\;,  
\end{equation}  
leading to $2^{M}$ pairwise orthogonal $2^{N-M}$-dimensional subspaces of   
the $N$-fermion Hilbert space ${\cal H}={\Bbb C}^{2^{N}}$.  

In fact, we may be even more general and assume some non-diagonal   
linear odd constraints obeying the algebra  
\begin{equation}  
\{\chi _{\alpha },\chi _{\beta }\}=w_{\alpha \beta }=w_{\beta \alpha }\;,
\quad w_{\alpha \beta }\in{\Bbb R}\;.
\end{equation}  
For simplicity we have chosen here self-adjoint odd second-class constraints.  
This system of constraints can easily be reduced to the above diagonal   
case. To be explicit, let $D\in SO(M)$ denote the orthogonal matrix   
which diagonalizes the symmetric matrix $W$,   
$(W)_{\alpha \beta }=w_{\alpha \beta }$. That is, we choose $D$ such   
that  
\begin{equation}  
(D^{T}WD)_{\alpha \beta }=v_{\alpha }\delta _{\alpha \beta }\;.  
\end{equation}  
Then we may define new constraints via 
$\chi '_{\alpha }=(D^{T})_{\alpha}{}^{\beta }\chi _{\beta }/\sqrt{v_\alpha}$ 
which are diagonal  
\begin{equation}  
\{\chi '_{\alpha },\chi '_{\beta }\}=\delta _{\alpha \beta }\;,
\end{equation}  
and can be treated as discussed above. Note that $v_\alpha>0$ as we are dealing
with second-class constraints.
  
In essence, the conclusion of this section is, that any set of linear
odd second-class constraints is reducible to the diagonal case and in  
turn can be incorporated into the path integral.

\subsection{Nonlinear odd constraints}  
Let us now consider odd constraints which are not linear in the fermion   
operators. Again we will begin our discussion with an elementary   
example which is an $N=4$ fermion system with constraints given by  
\begin{equation}  
\chi =f_{1}-f_{2}f_{3}f_{4}^{\dagger}\;,\quad  
\chi^{\dagger} =f^{\dagger}_{1}-f_{4}f^{\dagger}_{3}f^{\dagger}_{2}\;.  
\end{equation}  
Note that $\chi ^{2}=0=(\chi ^{\dagger})^{2}$ as before, however, the   
anti-commutator is no longer proportional to the identity. To be explicit,  
it is given by  
\begin{equation}  
\{\chi ,\chi ^{\dagger}\}=X  
\end{equation}  
where  
\begin{equation}  
X:={\bf 1}+f_{2}f_{2}^{\dagger}f_{3}f_{3}^{\dagger}f^{\dagger}_{4}f_{4}+  
f_{2}^{\dagger}f_{2}f_{3}^{\dagger}f_{3}f_{4}f^{\dagger}_{4}\;.
\end{equation}  
Note that $\mbox{spec}(X)=\{1,2\}$ and   
therefore its inverse is well-defined  
\begin{equation}  
X^{-1}={\bf 1}-\textstyle  
\frac{1}{2}f_{2}f_{2}^{\dagger}f_{3}f_{3}^{\dagger}f^{\dagger}_{4}f_{4}  
-\frac{1}{2}f_{2}^{\dagger}f_{2}f_{3}^{\dagger}f_{3}f_{4}f^{\dagger}_{4}\;.  
\end{equation}  
As in the linear case we cannot impose both conditions, case A and B in   
(\ref{55}), simultaneously. Hence, we again have to choose either case A or B.
Which will lead us to two orthogonal eight-dimensional subspaces of ${\cal   
H}={\Bbb C}^{16}$. Here, however, because of the non-linearity of the   
constraints, the projection operators are given by  
\begin{equation}  
\e_{A}=X^{-1}\chi \chi ^{\dagger}\;,\quad \e_{B}={\bf 1}-\e_{A}=X^{-1}\chi   
^{\dagger}\chi\;.  
\end{equation}  
Note that $[X,\chi ]=0=[X,\chi ^{\dagger}]$. In essence, because   
$X>0$ one simply replaces the original constraints by new ones, 
\begin{equation}  
\chi\to \chi'=\chi/\sqrt{X}\;, 
\end{equation}  
which by construction are ``linear'', i.e., constraints equivalent to linear, 
and can be treated as shown in the previous section. 
   
Obviously, this procedure can be generalized to a set of non-linear  
diagonal second-class constraints obeying  
\begin{equation}  
\{\chi_\alpha,\chi_\beta\}=0=\{\chi_\alpha^{\dagger},\chi_\beta^{\dagger}\}\;,
\quad\{\chi_\alpha,\chi_\beta^{\dagger}\}=X_\alpha\delta_{\alpha\beta}  
\end{equation}  
where $X_\alpha\geq 0$ does not vanish as $\chi _{\alpha }$ is assumed to be 
second class.  
Hence, we have $X_\alpha>0$ and therefore we may redefine the odd   
constraints $\chi_\alpha\to \chi'_\alpha=\chi_\alpha/\sqrt{X_\alpha}$  
which brings us back to the linear case discussed above.

\section{Application to Bose-Fermi systems}  
To complete our discussion we finally consider a system of $M$ bosons   
and $N$ fermions. The $M$ bosonic degrees of freedom are characterized by  
bosonic annihilation and creation operators $b_{i}$ and   
$b_{i}^{\dagger}$, respectively, which obey the standard commutation relations
\begin{equation}  
[b_{i},b_{j}]=0\;,\quad [b^{\dagger}_{i},b^{\dagger}_{j}]=0\;,\quad  
[b_{i},b^{\dagger}_{j}]=\delta _{ij}\;.  
\end{equation}  
These operators act on the $M$-boson Hilbert space   
$L^{2}({\Bbb R})\otimes\cdots\otimes L^{2}({\Bbb R})=  
L^{2}({\Bbb R}^{M})$. As in the case of fermions we will work in the   
(boson) coherent-state representation. These are eigenstates of the   
annihilation operators 
\begin{equation}  
b_{i}|z_{i}\rangle_{i}=z_{i}|z_{i}\rangle_{i}\;,\quad z_{i}\in {\Bbb C}\;, 
\quad |z_i\rangle_i\in L^2({\Bbb R})\;, 
\end{equation}  
and for its $M$-fold tensor product, which represents an $M$-boson state,  
we will use the notation   
$|\vec{z} \rangle=|z_{1}\rangle_{1}\otimes\cdots\otimes|z_{M}\rangle_{M}$.  
The total Hilbert space of the combined boson fermion system is thus ${\cal   
H}=L^{2}({\Bbb R}^{M})\otimes {\Bbb C}^{2^{N}}$ and the   
boson-fermion coherent states will be denoted by $|\vec{z} \Psi \rangle=  
|\vec{z} \rangle\otimes|\Psi \rangle$. The dynamics of such a system is   
defined by the Hamiltonian which we choose to  
\begin{equation}  
H:=\omega \left[\sum_{i=1}^{M}b_{i}^{\dagger}b_{i}   
+\sum_{i=1}^{N}f_{i}^{\dagger}f_{i}\right]\;,\quad \omega>0\;. 
\end{equation}
Note that for $M=N$ this Hamiltonian characterizes a supersymmetric quantum
system \cite{Nic76}.
The interaction of the bosons and fermions is introduced via the even  
first-class constraint 
\begin{equation}  
\Phi :=\sum_{i=1}^{M}b_{i}^{\dagger}b_{i}-  
\sum_{i=1}^{N}f_{i}^{\dagger}f_{i} - p\;,\quad p\in {\Bbb Z}\;,
\end{equation}  
which fixes the fermion number $N_{f}$ and the boson number $N_{b}$  
to obey the equality $N_{f}=N_{b}-p$.
  
As the spectrum of the constraint is integer we may   
use the integral representation (\ref{36}) for constructing the projection
operator. In this case the coherent-state matrix element for this 
operator reads 
\begin{equation}  
\begin{array}{l}
\langle\vec{z} ''\Psi ''|\e |\vec{z} '\Psi '\rangle\\[2mm]
\quad\displaystyle
={\cal N}\int_{0}^{2\pi }\frac{d\varphi }{2\pi }\,\E^{\I\varphi p}  
\exp\{\E^{-\I \varphi }\vec{z} ''^{*}\cdot\vec{z} '  
     +\E^{\I\varphi }\bar{\Psi}''\cdot \Psi '\}\;,  
\end{array}
\end{equation}  
where the normalization factor is given by  
\begin{equation}  
{\cal N}:=\exp\left\{-\frac{1}{2}\left[  
|\vec{z} ''|^{2}+|\vec{z} '|^{2}  
+\bar{\Psi }''\cdot\Psi ''+\bar{\Psi }'\cdot\Psi'\right]\right\}\;.  
\end{equation} 
Formally, the constrained propagator is represented by the path 
integral 
\begin{equation} 
\begin{array}{l}  
\langle\vec{z} ''\Psi ''|\E^{-\I tH}\e |\vec{z} '\Psi '\rangle= 
\displaystyle 
\int{\cal D}z^*{\cal D}z{\cal D}\bar{\Psi}{\cal D}\Psi{\cal D}C(\eta)\\[2mm]
\qquad\displaystyle\times
\exp\left\{\I\int_0^t d\tau\,L\right\}\;,\\[4mm] 
L:=\displaystyle\frac{\I}{2} 
(\vec{z}^*\cdot\dot{\vec{z}}-\dot{\vec{z}}^*\cdot\vec{z}+ 
\bar{\Psi}\cdot\dot{\Psi}-\dot{\bar{\Psi}}\cdot\Psi)\\[2mm]
\qquad\displaystyle
- \omega(\vec{z}^*\cdot\vec{z}+\bar{\Psi}\cdot\Psi)- 
\eta(\vec{z}^*\cdot\vec{z}-\bar{\Psi}\cdot\Psi-p)\;, 
\end{array} 
\end{equation} 
and explicit path integration leads to
\begin{equation}  
\begin{array}{l}  
\langle\vec{z} ''\Psi ''|\E^{-\I tH}\e |\vec{z} '\Psi '\rangle
=\displaystyle  
{\cal N }\int_{0}^{2\pi }\frac{d\varphi }{2\pi }\,\E^{\I\varphi p}\\[3mm]  
\qquad\displaystyle\times
\exp\left\{\E^{-\I (\omega t+\varphi)}\vec{z} ''^{*}\cdot\vec{z}'  
 +\E^{-\I (\omega t-\varphi)}\bar{\Psi}''\cdot \Psi '\right\}\\[2mm]  
\quad\displaystyle  
={\cal N}\sum_{m_{1}=0}^{\infty }\cdots \sum_{m_{M}=0}^{\infty } 
\sum_{n_{1}=0}^{1}\cdots \sum_{n_{N}=0}^{1} 
\delta _{\Sigma_N,\Sigma_M+p} \\[5mm]
\qquad\displaystyle\times
\frac{\E^{-\I \omega t(\Sigma_M+\Sigma_N)}}{m_{1}!\cdots m_{M}!}\\[4mm]  
\qquad\displaystyle\times  
\overline{  
(z_{1}'')^{m_{1}}\cdots (z_{M}'')^{m_{M}}  
(\psi_{1}'')^{n_{1}}\cdots (\psi_{N}'')^{n_{N}}}\\[2mm]
\qquad\displaystyle\times
(z_{1}')^{m_{1}}\cdots (z_{M}')^{m_{M}}  
(\psi_{1}')^{n_{1}}\cdots (\psi_{N}')^{n_{N}}  
\end{array}
\end{equation}  
where we have set $\Sigma_M:=m_1+\cdots+m_M$, $\Sigma_N:=n_1+\cdots+n_N$ and 
the overbar denotes an involution of the Grassmann algebra defined by
$\overline{c\psi _{1}\psi _{2}\cdots\psi _{N}}:=c^{*}\bar{\psi}_{N}\cdots  
\bar{\psi}_{2}\bar{\psi}_{1}$. 
  
\section{Conclusions} 
In this paper we have extended the bosonic coherent-state path-integral 
approach of constrained systems \cite{Klauder97} to those with fermionic 
degrees of freedom. As in the bosonic case we find that this approach does 
not involve any $\delta$-functionals of the constraints nor does it require 
any gauge fixing of first-class or elimination of variables for second-class 
constraints.
In addition we have shown that in the case of first-class constraints for
fermion systems it is sufficient to consider only those which have an even
Grassmann parity. In other words, for first-class constraints the Lagrange
multipliers are ordinary real-valued functions of time. There is no need to 
introduce either even or odd Grassmann-valued multipliers. In this respect
first-class constraints of fermion systems are not much different than those of
boson systems and can be incorporated in the path-integral approach in the same
way. This also applies to even second-class constraints. It is only in the case
of odd second-class constraints where Grassmann-valued Lagrange multipliers
may appear in the path-integral approach. 
For the cases of linear and non-linear diagonal second-class constraints 
we have 
been able to reduce the problem to the simpler case of linear diagonal
odd constraints which 
however does not
allow for a consistent quantum formulation. Here we have adopted a consistent
formulation by imposing only half (case A or B) of the second-class 
constraints. If one wants to avoid the appearance of Grassmann-valued Lagrange
multipliers at all then by virtue of relation (\ref{52}) one can choose for 
the projection operators 
$\e^{(\alpha)}_A$ and $\e^{(\alpha)}_B$ in Sect.\ 6.1 the simple integral 
representations
\begin{equation}
\e^{(\alpha)}_A=\int_0^{2\pi}\frac{d\eta}{2\pi}\,
\E^{-\I\eta\chi_\alpha^{\dagger}\chi_\alpha},\quad
\e^{(\alpha)}_B=\int_0^{2\pi}\frac{d\eta}{2\pi}\,
\E^{-\I\eta\chi_\alpha\chi_\alpha^{\dagger}}.
\end{equation}
This procedure in effect amounts to replacing the odd second-class
constraints $\chi_\alpha$ and $\chi_\alpha^{\dagger}$ by the even constraints 
$\Phi^{(\alpha)}_A:=\chi_\alpha^{\dagger}\chi_\alpha$ and 
$\Phi^{(\alpha)}_B:=\chi_\alpha\chi_\alpha^{\dagger}\;$, respectively. 
Note that from (\ref{68}) it immediately follows that for $\alpha\neq\beta$
\begin{equation}
[\Phi^{(\alpha)}_A,\Phi^{(\beta)}_A]=0\;,\quad
[\Phi^{(\alpha)}_A,\Phi^{(\beta)}_B]=0\;,\quad
[\Phi^{(\alpha)}_B,\Phi^{(\beta)}_B]=0\;.
\end{equation}
In other words, these even constraints are first class. So we finally conclude
that any odd first-class constraint and a wide range (linear and diagonal
non-linear) of odd second-class 
constraints appearing in fermion systems can be completely avoided within the 
approach presented in this paper.

\section*{Acknowledgement}
One of the authors (G.J.) would like to thank the Departments of Mathematics
and Physics of the University of Florida for their kind hospitality.
\setcounter{equation}{0}
\renewcommand{\theequation}{A.\arabic{equation}}
\section*{Appendix}
In this appendix we will present the explicit path-integral evaluations of
two examples discussed in the main text. The first one is for the system 
considered in Secton 4.1 whose formal path integral is given in (\ref{38}).
As measure for the Lagrange multipliers we choose
\begin{equation}
{\cal D}C(\eta)=\lim_{\varepsilon\to 0}\prod_{n=1}^{N}d\eta_n\,\delta(\eta_n)
\,\frac{d\xi}{2\pi}\langle\Psi_0|\E^{-\I\xi\Phi}|\Psi'\rangle
\end{equation}
which is normalized (in the $\eta$'s) and also introduces a projection operator
at $\tau=0$. Hence, the time-lattice path integral which we want to evaluate 
reads 
\begin{equation}
\begin{array}{l}
\displaystyle
\lim_{\varepsilon\to 0}\prod_{n=0}^{N-1}\int d\bar{\Psi}_n d\Psi_n
\int_0^{2\pi}\frac{d\xi}{2\pi}\\[2mm]
\displaystyle\qquad\times
\exp\left\{-\sum_{n=1}^N
\left[\frac{1}{2}\bar{\Psi}_n\cdot\Delta\Psi_n
      -\frac{1}{2}\Delta\bar{\Psi}_n\cdot\Psi_{n-1}\right]\right\}\\[2mm]
\displaystyle\qquad\times
\langle\Psi_0|\E^{-\I\xi\Phi}|\Psi'\rangle\;.
\end{array}
\end{equation}
Using the convolution formula
\begin{equation}
\begin{array}{l}
\displaystyle
\int d\bar{\Psi}_n d\Psi_n \,
\E^{-\bar{\Psi}_{n+1}\cdot\Delta\Psi_{n+1}/2
 +\Delta\bar{\Psi}_{n+1}\cdot\Psi_n/2}\\[2mm]
\qquad\displaystyle\times
\E^{-\bar{\Psi}_{n}\cdot\Delta\Psi_{n}/2
 +\Delta\bar{\Psi}_{n}\cdot\Psi_{n-1}/2}\\[2mm]
\displaystyle
\quad=
\E^{-\bar{\Psi}_{n+1}\cdot(\Psi_{n+1}-\Psi_{n-1})/2}
\E^{(\bar{\Psi}_{n+1}-\bar{\Psi}_{n-1})\cdot\Psi_{n-1}/2}\;,
\end{array}
\end{equation}
which follows from the completeness relation 
$\int d\bar{\Psi}_n d\Psi_n$ \linebreak
$\langle\Psi_{n+1}|\Psi_{n}\rangle$$\langle\Psi_{n}|\Psi_{n-1}\rangle=
\langle\Psi_{n+1}|\Psi_{n-1}\rangle$ and (\ref{10}), the path integral can be
reduced to
\begin{equation}
\begin{array}{l}
\displaystyle
\int d\bar{\Psi}_0 d\Psi_0\int_0^{2\pi}\frac{d\xi}{2\pi}\\[2mm]
\qquad\displaystyle\times
\exp\left\{-\frac{1}{2}\bar{\Psi}_{N}\cdot(\Psi_{N}-\Psi_{0})
+\frac{1}{2}(\bar{\Psi}_{N}-\bar{\Psi}_0)\cdot\Psi_0\right\}\\[2mm]
\qquad\displaystyle\times
\E^{\I\xi}
\langle\Psi_0|
\E^{-\I\xi(f_1^{\dagger}f_1+f_2^{\dagger}f_2)}
|\Psi'\rangle\;.
\end{array}
\end{equation}
The coherent-state matrix element appearing in the above expression is 
given by
\begin{equation}
\begin{array}{l}
\displaystyle
\langle\Psi_0|\E^{-\I\xi(f_1^{\dagger}f_1+f_2^{\dagger}f_2)}|\Psi'\rangle
=\E^{-\bar{\Psi}_0\cdot\Psi_0/2}\E^{-\bar{\Psi}'\cdot\Psi'/2}\\[2mm]
\qquad\displaystyle\times
\left[1+\E^{-\I\xi}\bar{\Psi}_0\cdot\Psi'
-\E^{-2\I\xi}\bar{\psi}_1\bar{\psi}_2\bar{\psi}'_1\bar{\psi}'_2\right],
\end{array}
\end{equation}
where we have used the notation $|\Psi'\rangle=|\psi'_1\rangle\otimes
|\psi'_2\rangle$ and $\langle\Psi_0|=\langle\psi_1|\otimes
\langle\psi_2|$. The remaining integrations are 
straightforward and lead to
\begin{equation}
\begin{array}{l}
\displaystyle
\int d\bar{\Psi}_0 d\Psi_0
\exp\left\{-\frac{1}{2}\bar{\Psi}''\cdot\Psi''
-\frac{1}{2}\bar{\Psi}'\cdot\Psi'\right\}\\[2mm]
\qquad\displaystyle\times
\exp\left\{(\bar{\Psi}''-\bar{\Psi}_0)\cdot\Psi_0\right\}
\bar{\Psi}_0\cdot\Psi'\\[2mm]
\displaystyle\quad
=\exp\left\{-\frac{1}{2}\bar{\Psi}''\cdot\Psi''
-\frac{1}{2}\bar{\Psi}'\cdot\Psi'\right\}\bar{\Psi}''\cdot\Psi'
\end{array}
\end{equation}
which is the result presented in (\ref{39}).
The evaluation of the path integral for the second example of first-class
constraints (see Section 4.2) is similar to that above.

As an example for an explicit path-integral calculation with second-class
constraints we choose case A of the linear odd constraint in Section 6.1.1. 
In this case the 
projection operator is given by $\e_A=|\theta\rangle\langle\theta|$ and the 
corresponding formal path integral (\ref{57}) reads in the time-lattice
formulation (\ref{45})
\begin{equation}
\begin{array}{l}
\displaystyle
\lim_{\varepsilon\to 0}\int\prod_{n=1}^{N-1}d\bar{\psi}_n d\psi_n\\[2mm]
\qquad\displaystyle\times
\exp\left\{\I\sum_{n=1}^{N}\left[
\frac{\I}{2}\bar{\psi}_n(\psi_n-\theta)
-\frac{\I}{2}(\bar{\psi}_n-\bar{\theta})\theta
\right.\right.\\[2mm]
\qquad\displaystyle\left.\left.
+\frac{\I}{2}\bar{\theta}(\theta-\psi_{n-1})
-\frac{\I}{2}(\bar{\theta}-\bar{\psi}_{n-1})\psi_{n-1}
-\varepsilon H(\bar{\theta},\theta)
\right]
\right\},
\end{array}
\end{equation}
where we have made use of the explicit form of the constrained
short-time propagator
\begin{equation}
\begin{array}{l}
\langle\psi_n | \e_A \E^{-\I\varepsilon H} \e_A |\psi_{n-1}\rangle\\[2mm]
\quad=\displaystyle\exp\left\{
-\frac{1}{2}\bar{\psi}_n(\psi_n-\theta)
+\frac{1}{2}(\bar{\psi}_n-\bar{\theta})\theta\right\}\\[2mm]
\qquad\times\displaystyle\exp\left\{
-\frac{1}{2}\bar{\theta}(\theta-\psi_{n-1})
+\frac{1}{2}(\bar{\theta}-\bar{\psi}_{n-1})\psi_{n-1}\right\}\\[2mm]
\qquad\times\displaystyle
\E^{-\I\varepsilon H(\bar{\theta},\theta)}\;.
\end{array}
\end{equation}
Rearranging the sum in the exponent the above path integral takes the
simple form
\begin{equation}
\begin{array}{l}
\displaystyle
\E^{-\bar{\psi}''(\psi''-\theta)/2}
\E^{(\bar{\psi}''-\bar{\theta})\theta/2}
\E^{-\bar{\theta}(\theta-\psi')/2}
\E^{(\bar{\theta}-\bar{\psi}')\psi'/2}
\E^{-\I t H(\bar{\theta},\theta)}\\
\qquad\times\displaystyle
\lim_{\varepsilon\to 0}
\prod_{n=1}^{N-1}\left[
\int d\bar{\psi}_n d\psi_n\E^{(\bar{\psi}_n-\bar{\theta})(\theta-\psi_n)}
\right].
\end{array}
\end{equation}
The remaining $N-1$ integration are easily evaluated providing $N-1$
factors of unity. Hence, we arrive at the result given in (58). The results 
(\ref{63}), (\ref{65}) and (\ref{66}) given in the main text are derived
in a similar fashion.



\begin{thebibliography}{References}
\bibitem{Comment}As an example of a constraint generating a non-compact group
let us suppose that the constraint $\Phi$ has a pure point spectrum including
zero which is the subspace of interest. Then the quantity
$\e=\int_{-\infty}^{\infty}d\xi\,\frac{\sin(\delta\xi)}{\pi\xi}\,\E^{\I\xi\Phi}$
leads us to a projection operator $\e=\e(-\delta<\Phi<\delta)$. 
For $\delta$ being
smaller than the gap to the closed discrete level then $\e$ is a projection 
operator onto the subspace where $\Phi=0$ as desired. 
Note that the above formula
does not make use of any spectral regularity and covers, e.g., the case of
$\Phi=N_1+\sqrt{2}\,N_2$, where $\mbox{spec}(N_1)=\mbox{spec}(N_2)=
\{0,1,2,\ldots\}$. For a
dense spectrum of $\Phi$, which would, e.g., arise for 
$\Phi=-N_1+\sqrt{2}\,N_2$,
we refer to the discussion in ref.\ \cite{Kla97b}.
\end{thebibliography}
\end{document}